# Rare Earth Ions Doped Mixed Crystals for Fast Quantum Computers with Optical Frequency Qubits

Vladimir Hizhnyakov[1], Vadim Boltrushko[1], Helle Kaasik[1], and Yurii Orlovskii[1,2]

1. Institute of Physics, University of Tartu, 1 W. Ostwald street, 50411 Tartu, Estonia
2. Prokhorov General Physics Institute, 38 Vavilov street, 119991 Moscow, Russia

E-mail: hizh@ut.ee

**Abstract**

The possibility of using mixed crystals highly doped with rare earth ions (REIs) as physical systems for creating fast quantum computers with a sampling time of $10^{-9}$ s is discussed. The electronic $^4f$ states of rare earth ions with small values of the diagonal elements of the Judd-Ofelt matrix $U^{(2)}$ are proposed as optical frequency qubit levels. CNOT and other conditional gate operations are performed by exciting the rare earth ion into the $^4f$ state with a large diagonal element of $U^{(2)}$, causing a Stark blockade. It is found that the main interaction responsible for this blockade is the quadrupole-quadrupole interaction. The large inhomogeneous broadening of the frequencies of the electronic transitions in mixed crystals and the weak interaction of $^4f$ electrons with phonons make it possible to achieve a high computation rate and a long decoherence time of the qubits. An ensemble of closest REIs is described that can act as an OQC instance; the frequencies of the corresponding qubits can be found using the spectral hole burning method.

## 1. Introduction

In recent years, many researchers working on the use of rare earth ions (REI) in crystals for the fabrication of optical quantum computers (OQC) have tied their expectations to the combined use of hyperfine and optical transitions in REI with hyperfine levels used as qubits (see, e.g. [1-15]). These systems were found to meet the known criteria of quantum computers (DiVincenzo's criteria) [16]. The qubits considered in [1-15] are very weakly dependent on the environment. Such qubits interact extremely weakly with each other and can be considered independent, as required for quantum computers. At the same time, the electronic transitions in REI are more dependent on the environment, especially on the electronic states of other neighbouring REI. This dependence can be used to implement CNOT and other conditional gate operations with qubits, using a sequence of optical pulses that temporarily transfer the corresponding REI to another electronic state [1-15]. For this purpose, it has been proposed (see papers [1-15]) to use the phenomenon of the dipole (Stark) blockade [17-20]. The optical excitation of REI changes the surrounding static field. The latter in turn changes the transition frequencies in other optical centers, thus breaking the resonance for their optical transitions.

One-qubit operations on qubits with hyperfine levels require two optical pi-pulses [1-15]. These pulses must be long enough to have a small spectral width, which must be much less than the qubit frequency. This is due to the fact that short pulses with a duration comparable to the reciprocal frequency of such qubits have a spectral width comparable to their frequency and, therefore, cannot lead to the creation of purely basic qubit states. Therefore, OQCs with high fidelity and scalability using qubits with hyperfine levels should be relatively slow, with typical sampling times of one microsecond or longer. In this communication we discuss the possibility of using the REI to create a fast (~GHz) OQC. This becomes possible if different electronic 4f REI levels are taken as qubits. These qubits have optical frequencies. In this case, the spectral width of a nanosecond pulse is incomparably smaller than the qubit frequencies. This allows pi-pulses to be used to create purely basic states of qubits. Therefore, in this case, the OQC can be obtained with high fidelity. Another difference between OQC with optical frequency qubits is that, single qubit gate operations, in contrast to those described in Refs. [1-15], can be performed with single light pulses. At the same time, CNOT and other conditional gate operations on qubits can be performed in the same way, using multiple pulses and a Stark blockade.

Previously it was considered that the main interaction responsible for the Stark blockade is the dipole-dipole interaction [1-15, 19, 20]. Therefore, the term "dipole blockade" was used to refer to this phenomenon. However, according to our considerations, this is true only at large distances between REIs ($\gtrsim$ 10 lattice constants); due to the small oscillator strength of dipole transitions between 4f states, the quadrupole-quadrupole and dipole-quadrupole interactions are more important at shorter distances. (The predominance of these interactions of REIs at small distances was observed for incoherent nanosecond energy transfer in [21].) Furthermore, we found that the frequency shift of the electronic transition in one REI, caused by the electronic excitation of another REI, leading to a Stark blockade, is determined by the diagonal matrix elements of the Judd-Ofelt matrix $U^{(2)}$ [22-24]. These matrix elements have been calculated in a series of works for practically all $^4f$ levels of REI (see, e.g. [25, 26]). These calculations show a remarkable property: for all REI the elements of the $U^{(2)}$ matrix have extremely scattered values, which differ by many orders of magnitude. This important property allows to successfully solve the dilemma of independence of qubits and their control by other qubits. Indeed, ***for OQC, $^4f$ levels of REI with small diagonal elements of the matrix $U^{(2)}$ can be used as optical frequency qubits; and the $^4f$ levels with large diagonal elements of the matrix $U^{(2)}$ can be used as auxiliary levels to implement CNOT and other conditional gate operations.***

Note that the use of electronic $^4f$ REI states for OQC qubits has been recently discussed in [27-29], but without noting the importance of using states with small and large diagonal elements of the Judd-Ofelt matrix $U^{(2)}$.

In this communication, we discuss the possibility of creating fast (GHz) OQC using mixed crystals doped with trivalent rare-earth ions. As specific systems, we consider mixed $La_{1-x}Y_xF_3$ crystals and their analogues. The mixed crystals under consideration have large differences in the strengths of the crystal field at different points of the crystal lattice. These large differences lead to a large inhomogeneous linewidth ($\Gamma_{inh}$ ~ THz) of zero phonon lines (ZPLs) in optical spectra [30]. This is important for fast quantum computers, since a short sampling time $t_0$ in the order of nanoseconds requires the use of laser pulses with a sufficiently wide spectrum of width $\Gamma_L = 1/t_0$ ~ GHz. We take into account that in the case of $\Gamma_{inh} \gg \Gamma_L$ and $\Gamma_L > \Gamma_h$ a large number of qubits can be individually addressed with single laser pulses of different frequencies ($\Gamma_h$ is the homogeneous width of ZPL). This is different from OQC with hyperfine qubit levels where two pulses are used to address qubits individually.

Quantum computing is possible if the dephasing time of the qubit (transverse relaxation time) $T_2 \sim \Gamma_h^{-1}$ is long - significantly longer than the sampling time $\sim \Gamma_L^{-1}$. In the systems under consideration, this condition can be met due to the weak interaction of the $^4f$ electrons of REI with the environment and the structural features of the considered mixed crystals. We take into account that in mixed $La_{1-x}Y_xF_3$ crystals, as well as in single (perfect) crystals, the coordination numbers are conserved. In this way, they differ from the mixed crystals in which this conservation law is violated. This violation has an important consequence, namely low-energy excitations (tunnel systems and pseudo-local modes) are generated, which leads to a strong increase in the rate of decoherence [31-40]. In the proposed mixed crystals, these extra excitations are absent (or *almost* absent). Therefore, as with single crystals, (in addition to the decay rate), $\Gamma_h$ is determined by the phonon Raman mechanism, which gives at low temperatures $\Gamma_h \propto T^7$ [41, 42]. Therefore, it is expected that at sufficiently low temperatures $\Gamma_h \sim T_2^{-1}$ is determined by the decay rate of the qubit levels $\gamma$.

In many cases $\gamma \lesssim 10^6 \text{ s}^{-1}$, and $T_2 \gtrsim$ than one microsecond [43]. For example, the transverse relaxation time of the $^4I_{13/2} \rightarrow {}^4I_{15/2}$ transition of $Er^3$ in a $Eu^{3+}$: $Y_2SiO_5$ crystal at 1.5 K is $T_2 = 3.3$ µs, increasing in the $Er^{3+}$: $Y_2O_3$ crystal to $T_2 = 18$ µs. [44]. Moreover, in some cases one has even larger $T_2$. For example, the upper estimate of the $T_2$ in crystals without an external magnetic field for metastable levels, measured at the $^7F_0 \rightarrow {}^5D_0$ transition of $Eu^{3+}$ in a $Eu^{3+}$: $Y_2SiO_5$ crystal at 1.4 K, is $T_2 = 1.5$ ms [45] which is three order of magnitudes longer than one microsecond. Consequently, in REI crystals one can get at low temperatures $\Gamma_{inh}/\Gamma_h > 10^5$, which is sufficient for various quantum calculations. In crystals with La and other REI as cations, this condition is normally met for the temperature of liquid helium. However, one would expect that when using mixed crystals with lighter cations (for example, mixed $BF_3+AlF_3$ crystals), this condition is also fulfilled, and corresponding quantum computers could operate at higher temperatures.

It is important to note that only the qubit levels of the fast OQC should have a long decoherence time $1/\pi\Gamma_h \gtrsim 10^{-7}$ s; auxiliary levels used for CNOT and other conditional gate operations may have much shorter coherence times in the order of several sampling times ($\sim 10^{-8} - 10^{-9}$ s). This is the advantage of the considered fast OQC compared to the OQC considered in [1-15], where all used 4f levels should have a long coherence time.

To make an OQC, it is necessary to use a microcrystal that provides the same light pulse intensity for all qubits, which is necessary to achieve high fidelity and scalability. In this microcrystal, a group of $N \sim 100$ closely spaced centers (qubits) must be separated, in which the operations of the CNOT gates can be performed for all pairs of centers. To achieve this, one can use the method of spectral hole burning. (This method was proposed to be used for OQC in [1, 3, 8], although in a different way.) To apply this method, first, in the microcrystal under consideration, an optical center is excited with a light pulse. Such excitation due to the Stark blockade leads to the appearance of hole-antihole pairs in the absorption spectrum. The larger the spectral distance between the hole and the antihole in the hole-antihole pair, the stronger the interaction with the initially excited center and the closer the distance to them. Then one should take $N'$ spectral holes having the largest distances between holes and antiholes. The frequencies of these holes will give the desired frequencies of the centers (qubits) closest to each other, forming the working ensemble of ions. Note that the concentration of the initially excited centers $c\Gamma_L/\Gamma_{inh}$ is very low ($c$ is the concentration of working REI ions). Therefore, a microcrystal contains only a small number $k$ of such ensembles. Using the subsequent hole burning operations, one can select one of them to act as an OQC with $N = N'/k$ qubits.

These issues are discussed in more detail below.

## 2. Single pulse gate operations

Considered quantum computers with electronic $^4f$ levels of REI as qubits can operate with the help of a sequence of resonant light pulses. The qubit transformation caused by such pulses is described by the operator [46, 47]

$$V(\Theta,\varphi) = I\cos\Theta/2 + i(\sigma_x\sin\varphi + \sigma_y\cos\varphi)\sin\Theta/2, \tag{1}$$

where $\sigma_x$ and $\sigma_y$ are Pauli matrices, $I$ is $2\times 2$ unit matrix, $\Theta = e|E|\langle 0|r|1\rangle/\hbar\Gamma_L$ is given by the product of the interaction energy $Ee\langle 0|r|1\rangle$ of the laser pulse with the qubit divided by $\hbar$ and the duration of the pulse $\Gamma_L^{-1}$; $e\langle 0|r|1\rangle$ is the dipole matrix element of the transition $|0\rangle \to |1\rangle$ in a qubit; $e$ is the elementary charge; $E$ is the electric field strength of the laser pulse, $\varphi$ is its phase.

Let us estimate the power density of laser radiation required for gate operations with qubits. Taking into account that the electro-dipole mechanism is the main contributor to $\gamma_0$ of the most radiative transitions of REI in crystals [19, 20]. In this case

$$|\langle 0|r|1\rangle| = \sqrt{3\gamma_0/4\alpha ck^3}, \tag{2}$$

where $\alpha = e^2/\hbar c = 1/137$ is the fine structure constant, $\gamma_0$ is the radiative decay rate, $k = \omega n/c$ is the wave number of light with frequency $\omega$, $n$ is refractive index. With this relation, we obtain the following field strength of the pulse for $\Theta = \pi$:

$$|E| = 2\pi\,\Gamma_L\sqrt{\hbar k^3/3\gamma_0}. \tag{3}$$

The light intensity of this pulse is $I = E^2/Z$, its energy equals $\mathrm{E}_L = E^2 S/\Gamma_L Z$, where $Z = 376.7$ ohm is the impedance of free space, $S$ is the cross-section of the beam of the laser pulse. The rate $\gamma_0$ can be taken from the experiment. For example, in the case of $^3H_4 \to {}^3P_0$ electronic transition in LaF$_3$:Pr$^{3+}$ ($\omega = 20469\,\mathrm{cm}^{-1}$, $n = 1.6$, $\gamma_0 \approx 1.8\cdot 10^4$ sec$^{-1}$; see [48, 49] and the references therein), according to Eq. (3), the $\pi$-pulse of ~1 ns duration (with $\Gamma_L \sim \mathrm{GHz}$) should have the field strength $E \sim 3\cdot 10^4$ V/cm. This corresponds to the power density $\sim 2$ MW/cm$^2$, which can be easily achieved with modern laser systems and does not lead to radiation damage to the crystals (the latter occurs in pulses with a power six orders of magnitude greater).

Note that in the case of microcrystals, the field strength is the same with high precision for all REIs, which makes it easier to achieve high fidelity of OQC.

## 3. Stark blockade

To perform control gate operations, the corresponding REI must interact with each other [46, 47]. Here we consider the interaction that occurs due to the fact that a change in the electronic state of the REI leads to a change in the static crystal field around it. The change of the field in turn leads to a Stark shift in the frequencies of electronic transitions of other REI [17-20]. The frequencies of some transitions can change so strongly that they are no longer in resonance with the frequencies of the transitions of the unexcited first center. Such a loss of resonance has been described in [17-19]. In [1-9] it was shown that this phenomenon, called dipole (Stark) blockade, can be used to implement CNOT and other control gate operations.

It is generally accepted that the Stark blockade is caused by the dipole-dipole interaction of the centers [1-15, 17-20]; therefore, it was described in [20] as the dipole blockade. This interaction is fundamental in the case of centers that are located at large distances from each other. However, we are interested in the case, where the centers are located at intermediate distances of several to ten lattice constants. In the case of REI, the dipole-dipole interaction is rather weak due to the low oscillator strength of the $|{}^4f\rangle \leftrightarrow |{}^4f'\rangle$ transitions. Therefore, for such centers, it is necessary to consider the blockade caused by their quadrupole- quadrupole interaction. For intermediate distances, the dipole-quadrupole interaction can also matter.

In order to find the change of the frequency of an electronic transition in a center caused by the static field, which is created when another center changes its electronic state, it is necessary to expand the Coulomb interaction of the optical electrons of these centers in terms of coordinates $\vec{r}$ of these electrons. The first-order terms will describe the dipole-dipole interaction, and the second-order terms will describe the quadrupole-quadrupole interaction. In the case of the dipole-dipole interaction, this gives the following change in the frequency of the electronic transition $|j\rangle_{m1} \leftrightarrow |j'\rangle_{m1}$ at the center $m_1$ due to the excitation $|j\rangle_{m2} \leftrightarrow |j'\rangle_{m2}$ of the center $m_2$

$$\delta_{jj'}^{(d)}\Big|_{m1,m2} = \frac{e^2}{\varepsilon_0 R^3}\left(-2\langle \bar{x}_{jj'}^s\rangle_{m1}\langle \bar{x}_{jj'}^s\rangle_{m2} + \langle \bar{y}_{jj'}^s\rangle_{m1}\langle \bar{y}_{jj'}^s\rangle_{m2} + \langle \bar{z}_{jj'}^s\rangle_{m1}\langle \bar{z}_{jj'}^s\rangle_{m2}\right). \tag{4}$$

Here $\langle \bar{r}_{jj'}^s\rangle_m = \langle j'|r_\alpha^s|j'\rangle_m - \langle j|r_\alpha^s|j\rangle_m$ is the change of the mean value of $r_\alpha^s$ of the center $m$ at the electronic transition; $(r_x, r_y, r_z) \equiv (x, y, z)$, the $x$ axis is taken in the direction of the vector $\vec{R}$.

According to the Judd and Ofelt theory [22, 23], the matrix element $|\langle 0|r_\alpha|1\rangle|$ is determined by the admixture of $^5d$ states of opposite parity caused by the low-symmetric crystal field of the nearest ligands. For all cases (except for the nearest ion), this admixture is given only by the first power-law corrections with respect to the coordinates of the $^4f$ electrons [22, 23]. Therefore, the dipole matrix element $|\langle 0|r_\alpha|1\rangle|$ for the corrected $^4f$ states is actually determined [22, 23] by the matrix element of $r^2$ for $^4f$ states of aqua ions $|j^{(0)}\rangle$, the second power of $r$ being derived from the linear ones with respect to the $r$ corrections of the wave functions of $^4f$ levels caused by the weak asymmetric crystal field. The contributions of these terms are described by the $U_{jj'}^{(2)}$ matrix elements of the Judd and Ofelt theory [22, 23].

In the case of quadrupole-quadrupole interaction, the frequency change equals to

$$\delta_{jj'}^{(q)}\Big|_{m1,m2} = \frac{3e^2}{4\varepsilon_0 R^5}\Big[\left(\langle \bar{r}_{jj'}^2\rangle_{m1} - 5\langle \bar{x}_{jj'}^2\rangle_{m1}\right)\left(\langle \bar{r}_{jj'}^2\rangle_{m2} - 5\langle \bar{x}_{jj'}^2\rangle_{m2}\right) - \\ 8\langle \bar{x}_{jj'}^2\rangle_{m1}\langle \bar{x}_{jj'}^2\rangle_{m2} + 2\langle \bar{y}_{jj'}^2\rangle_{m1}\langle \bar{y}_{jj'}^2\rangle_{m2} + 2\langle \bar{z}_{jj'}^2\rangle_{m1}\langle \bar{z}_{jj'}^2\rangle_{m2}\Big] \tag{5}$$

The matrix elements can be calculated using wave functions $|j^{(0)}\rangle$ of aqua ions. In this approximation, the matrix elements are determined by the matrix elements $U_{jj'}^{(2)}$.

For nearest REI it is necessary to consider also higher (up to fifth) order corrections of $^4f$-wave functions [19, 20]. As a result, the corrected wave functions are described by the matrices

$$U_{jj'}^{(k)} = \left\langle j^{(0)}\left|(r/r_0)^k\right|j'^{(0)}\right\rangle, \tag{6}$$

with $k = 2, 4, 6$ [22, 23]. The results of calculations of these matrices are given in Refs. [25, 26].

To estimate $\delta^{(d)}$, we put it in the form

$$\delta^{(d)} \sim \gamma_0 \varepsilon_0^{-1}\left(\tilde{U}_{(0,1)}^{(2)}/U_{01}\right)^2 (kR)^{-3} \tag{7}$$

where $\tilde{U}^{(2)}_{(0,1)}$ is the largest of $\left|U^{(2)}_{00}\right|$ and $\left|U^{(2)}_{11}\right|$, $U_{01}$ is the matrix element $U^{(n)}_{01}$, $n = 2,4,6$ giving the largest contribution to $\gamma_0$. According to Refs. [22-26], the absolute values of these matrix elements in most cases are in the range from 1 to 0.001, and change with the change of states. The rates of radiative transitions between $^4f$ states are usually in the range of $\gamma_0 \sim 10^4$ sec$^{-1}$. Taking $(\tilde{U}^{(2)}_{(0,1)}/U^{(2)}_{00})^2 \sim 1$, and $\varepsilon_0 \sim 10$, we get $\delta^{(d)} \sim 100(a/R)^3$ GHz (here $a \sim 4$Å is the lattice constant). This value agrees with the estimation for this quantity given in work [41].

To estimate $\delta^{(q)}$, consider two centers at distance $R$ in the $x$-direction. We get

$$\delta^{(q)} \sim 25\left(U^{(2)}_{11} - U^{(2)}_{00}\right)^2 \bar{\omega}_0 \, r_0^5 / \varepsilon_0 R^5 \qquad (8)$$

where $\bar{\omega}_0 \sim 3$ PHz is the mean frequency of $^4f$ - $^4f$ transitions. Taking $\left(U^{(2)}_{11} - U^{(2)}_{00}\right)^2 \sim 0.1$, $r_0^2 \sim 0.1 a^2$ and $\varepsilon_0 \sim 10$, we get $\delta^{(q)} \sim 50(a/R)^5$ THz.

For REI optical centers being at an intermediate distance $R \sim 5a$ one gets $\delta^{(d)} \sim 5$ GHz and $\delta^{(q)} \sim 30$ GHz. Consequently, *for the intermediate and small distances between REI, which are important for the OQC under consideration, the main interaction leading to the Stark blockade is the quadrupole-quadrupole interaction.*

One can see that both $\delta^{(d)}$ and $\delta^{(q)}$ are determined by the Judd-Ofelt parameters $U^{(2)}_{jj'}$. The same is true for the delta $\delta$ given by the dipole-quadrupole interaction. Therefore, for all distances the condition $\delta > 1$ GHz of the Stark blockade is most easily met for levels with large $\tilde{U}^{(2)}_{(0,1)}$.

The centers under consideration have non-zero spin which also contributes to $\delta$. In the case of Kramers REI's with half-integer spin, the strongest spin-spin interaction at short distances is the exchange interaction $J$ [50, 51]. For example, for nearest pair centers of Nd$^{3+}$ this interaction is of the order of several cm$^{-1}$, and it is responsible for the singlet-triplet splitting $2J \sim 2-5$ cm$^{-1}$ of the ground state [50-52]. With increasing distance, this interaction decreases exponentially, and for $R > 3a$ it can be neglected. The remaining spin-spin interaction is of magnetic dipole-dipole origin and decreases with increasing distance as $R^{-3}$. For the nearest neighbours it is of the order of $\sim 10^{-2} - 10^{-1}$ cm$^{-1}$ being comparable to or smaller than $\delta^{(d)}$ [51].

Above it is supposed that qubits can be addressed independently. This is the case when the interaction between REIs does not cause excitation delocalization. Usually the energy exchange takes place mostly for nearby centers. To fulfil the corresponding condition, we take into account that according to the motional narrowing effect, the transfer of excitation between two REI's due to the exchange interaction takes takes place with the rate $\sim \Gamma^2_{exch}/\left(\Gamma^2_{exch} + \Delta^2\right)^{1/2}$, where $\Gamma_{exch}$ is the strength of the resonance exchange interaction, $\Delta$ is the difference of the transition frequencies in the REI. A rough estimation gives $\Gamma_{exch} \sim \delta$. In the case of large inhomogeneous broadening $\Gamma_{inh} \sim$ THz, the condition $\Delta \gg \delta$ is fulfilled for all concentrations of REI's, which means that delocalization of excitations can be neglected. This conclusion holds also for the Förster-type excitation transfer occurring with emission or/and absorption of phonons. Due to the weak coupling of $^4f$ electrons with phonons, only one-phonon processes are significant at low temperature [53]. These transitions have the rate $\Gamma_F \sim \Gamma^2_{exch} \kappa^2 (\bar{n}_\delta + 1)/|\Delta|$ [54], where $\kappa \ll 1$ is the dimensionless constant of electron-phonon interaction, $\bar{n}_\delta = (e^{\hbar\delta/k_B T} - 1)^{-1}$ is the Planck phonon population factor.

The exchange rate for this mechanism depends on the distance as $(r_0/R)^{10}$, and it is important only for nearby REI's.

Note that some REI's form cooperative states. In this case one usually speaks of pair, triple, etc. centers [51,52,55]. In the case under consideration ($\Delta \gg \delta$), the number of such centers is low.

## 4. Gate operations with two and more qubits

Let us look at use of the Stark blockade in REI for CNOT and other control gate operations. Recall that CNOT corresponds to performing the NOT gate operation in the first (target) qubit $m_1$ if the second (control) qubit $m_2$ is in the state $|1\rangle_{m2}$. However, this operation is not performed, if the second qubit is in the state $|0\rangle_{m2}$. It turns out that both, single-qubit and control gate operations, are possible for the $^4f$ levels of REI, since these ions have a set of $^4f$ levels with very different values of $\left|U_{jj'}^{(2)}\right|$ parameters that determine the strength of the interaction of REI. In particular, REI's have levels with small and large values of $\left|U_{jj}^{(2)}\right|$. The levels with large $\left|U_{jj}^{(2)}\right|$ can be used as the auxiliary levels $|1'\rangle$ to perform the control for the CNOT operation, since a large $\left|U_{jj}^{(2)}\right|$ simplifies the fulfilment of the condition of the Stark blockade

$$\delta_{1',j}\big|_{m1,m2} > \Gamma_L, \Gamma_h. \tag{9}$$

In this case, a five-pulse scheme (see Figure 1), similar to that proposed in [1], can be used to implement the CNOT gate operation, but without using hyperfine states.

To implement CNOT gate operations in a system of $N$ qubits, when using the microcrystals under consideration, it is necessary to apply the method of spectral hole burning and gain saturation in doped solids. In this method, the excitation of a center due to the Stark blockade, leads to the appearance of hole-antihole pairs in the absorption spectrum. For the considered mixed crystals, due to the large inhomogeneous width of the ZPL, it is possible to distinguish a large number $N'$ of hole-antihole pairs giving the frequencies of closely located optical centers. The greater the spectral distance between the hole and the antihole in the hole-antihole pair, the stronger the interaction with the initially excited optical centers, causing the Stark blockade, and the closer the distance to them. The found frequencies of qubits correspond to a number of disjunctive, independent ensembles of closely spaced centers (qubits). Each of these ensembles of qubits may act as an instance of a quantum computer. The concentration of the initially excited centers $c\Gamma_L/\Gamma_{inh}$ is very low. Therefore, the microcrystal under consideration contains only a small number $k$ of such ensembles of centers. Using the subsequent hole burning operations, one can select one of them to act as an OQC with $N = N'/k$ qubits.

As an example, consider a mixed crystal doped with REIs with a concentration $c = 0.1$. Let us assume that $\Gamma_{inh} \sim 1$ THz and $\Gamma_h \ll 1$ GHz. In the considered case of excitation of REIs by a laser pulse with a spectral width of $\Gamma_L \sim 1$ GHz, the concentration of being excited REIs is $c\Gamma_L/\Gamma_{inh}$ giving $\sim 22\,a$ for the mean distance of these REIs. Consider an ensemble of $N$=50 REIs closest to the excited one, which can act as an OQC instance. The mean size of this ensemble is $(N/c)^{1/3} a = 7.9\,a$. According to Eq. (8), the weakest interaction between REIs in this ensemble of ions $\delta^{(q)} \sim 3$GHz exceeds $\Gamma_L$. Therefore, in this case, fast CNOT gate operations can indeed be successfully performed for all pairs of $N$=50 working qubits.

To perform CNOT in OQC under consideration, first, a $\pi$-pulse with the frequency $\omega_{0,1'}\big|_{m2}$ is applied to the optical center $m_2$ of the control qubit (see Figure 1, the first pulse). Then this center will either reach the auxiliary level or not, depending on whether it was in the state $|0\rangle_{m2}$ or in the state $|1\rangle_{m2}$. In turn, the frequency $\omega_{1'}\big|_{m1}$ of the auxiliary level $|1'\rangle_{m1}$ of the center $m_1$ of the target qubit will change to $\omega_{1'}\big|_{m1,m2} = \omega_{1'}\big|_{m1} + \delta\big|_{m1,m2}$ (and will go out of resonance) or not, depending on whether the second (control) qubit was in the state $|0\rangle_{m2}$ or in the state $|1\rangle_{m2}$. Therefore, if we use three-pulse transitions $|0\rangle_{m1} \leftrightarrow |1'\rangle_{m1}$, $|1'\rangle_{m1} \leftrightarrow |1\rangle_{m1}$ and $|0\rangle_{m1} \leftrightarrow |1'\rangle_{m1}$ to perform the NOT gate operation on the target qubit $m_1$ (see the second, third, and fourth pulses in Figure 1), then the success of this operation depends on the initial state of the control qubit: if it is in the state $|0\rangle_{m2}$, then the transitions $|0\rangle_{m1} \leftrightarrow |1'\rangle_{m1}$, $|1'\rangle_{m1} \leftrightarrow |1\rangle_{m1}$ and $|0\rangle_{m1} \leftrightarrow |1'\rangle_{m1}$ in the target qubit will not occur due to the Stark blockade of the state $|1'\rangle_{m1}$, and the NOT gate operation in this qubit will not be performed. However, if the control qubit is in the state $|1\rangle_{m2}$, then the transitions $|0\rangle_{m1} \leftrightarrow |1'\rangle_{m1}$, $|1'\rangle_{m1} \leftrightarrow |1\rangle_{m1}$ and $|0\rangle_{m1} \leftrightarrow |1'\rangle_{m1}$ occur in the target cubit (see pulses 2, 3, and 4 in Figure 1), and the NOT operation is performed in the target qubit. To complete the CNOT operation, the last $\pi$-pulse is applied to the center $m_2$ of the control qubit (see pulse 5 in Figure 1), initiating the transition $|1'\rangle_{m2} \to |0\rangle_{m2}$ to return it to its original state.

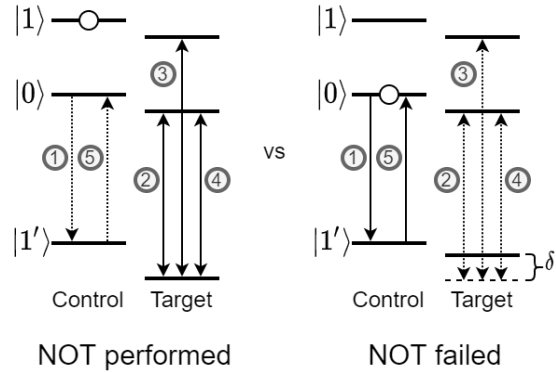

Fig. 1. The scheme of CNOT operation. The arrows indicate the excitation by $\pi$-pulses of light; the numbers next to them are the numbers of the pulses in the sequence. The circles indicate the initially occupied levels of the control qubits. $\delta$ is the shift of the auxiliary level in the target qubit due to a change of state of the control qubit. The relative position of the levels can be different.

We present here several examples of the implementation of this scheme, listed in Table 1.

Table 1. REI's and their electronic levels suitable for implementation of CNOT quantum gate

| state | $Pr^{3+}$ | $Er^{3+}$ | $Tm^{3+}$ | $Tm^{3+}$ | $Tm^{3+}$ | $Tm^{3+}$ |
|---|---|---|---|---|---|---|
| $|0\rangle$ | $|{}^1G_4\rangle$ | $|{}^4I_{9/2}\rangle$ | $|{}^3F_4\rangle$ | $|{}^3H_4\rangle$ | $|{}^3F_4\rangle$ | $|{}^3F_4\rangle$ |
| $|1\rangle$ | $|{}^3P_0\rangle$ | $|{}^4S_{3/2}\rangle$ | $|{}^1D_2\rangle$ | $|{}^1D_2\rangle$ | $|{}^1D_2\rangle$ | $|{}^3H_4\rangle$ |
| $|1'\rangle$ | $|{}^3H_4\rangle$ | $|{}^4I_{15/2}\rangle$ | $|{}^3H_6\rangle$ | $|{}^1I_6\rangle$ | $|{}^1I_6\rangle$ | $|{}^1I_6\rangle$ |

First, we consider the use of $Pr^{3+}$ and $Er^{3+}$ ions for OQC. The possible options for using the ${}^4f$ levels of these ions for qubits and auxiliary levels are shown in Table 1. These options have the common feature that the state $|1'\rangle$ has a lower energy than the states $|0\rangle$ and $|1\rangle$. The scheme of CNOT gate operation for these cases is shown in Figure 1. Transitions between all selected levels in these ions are allowed: see values of non-diagonal $U^{(2)}_{jj'}$- parameters in Table 2. The values of $U^{(2)}_{ff'}$ parameters are taken from [26] and from unpublished results of A. Kornienko, which are very close to the values given in [25]. The calculations begin with the preparation of qubits in the initial states $|0\rangle_m$ by exciting the transition $|1'\rangle_m \to |0\rangle_m$ wiFth the help of a $\pi$-pulse.

Table 2. Energies of electronic levels $E$ (cm$^{-1}$), their lifetimes, and squares of reduced matrix elements of electronic transitions $U^{(k)}$.

| ion | state | level | $E$, cm$^{-1}$ | $\tau$, $\mu$s | $|U^{(2)}|^2$ | | | $|U^{(4)}|^2$ | | | $|U^{(6)}|^2$ | | |
|---|---|---|---|---|---|---|---|---|---|---|---|---|---|
| | | | | | $|0\rangle$ | | | $|1\rangle$ | | | $|1'\rangle$ | | |
| Pr$^{3+}$ | $|0\rangle$ | $|^1G_4\rangle$ | 9640 | 14 | **0** | 0 | 0 | 0 | .056 | 0 | .012 | .072 | .027 |
| | $|1\rangle$ | $|^3P_0\rangle$ | 20469 | 55 | 0 | .056 | 0 | **0** | 0 | 0 | 0 | .173 | 0 |
| | $|1'\rangle$ | $|^3H_4\rangle$ | 0 | $\infty$ | .012 | .072 | .027 | 0 | .173 | 0 | **.779** | 0 | 0 |
| Er$^{3+}$ | $|0\rangle$ | $|^4I_{9/2}\rangle$ | 12272 | 133 | **.002** | 0 | 0 | 0 | .079 | .254 | 0 | .173 | .010 |
| | $|1\rangle$ | $|^4S_{3/2}\rangle$ | 18353 | 923 | 0 | .079 | .254 | **.037** | **0** | 0 | 0 | 0 | .223 |
| | $|1'\rangle$ | $|^4I_{15/2}\rangle$ | 0 | $\infty$ | 0 | .173 | .010 | 0 | 0 | .223 | **.247** | 0 | 0 |

For OQC, Pr$^3$ ions are used with the ground state $^3H_4$ as the auxiliary states and the states $^1G_4$ (in Pr$^{3+}$: YLiF$_4$ $\tau$ = 14 µs) [56] and $^3P_0$ (in Pr$^{3+}$:LaF$_3$ $\tau$ = 55 µs) [48, 49] as qubit states. The calculations begin with preparation of states $|0\rangle_m$ using the $\pi$-puls excitation of the $|3H_4\rangle \to |^1G_4\rangle$ transition. In case of Er$^{3+}$ ions, one can use the ground state $^4I_{15/2}$ as the auxiliary level and states $^4I_{9/2}$ (in Er$^{3+}$:LaF$_3$ $\tau$= 133 µs) [56] and $^4S_{3/2}$ (in Er$^{3+}$:LaF$_3$ $\tau$ = 923 µs) [57] as qubit levels (see parameters in Table 2). Instead of $^4S_{3/2}$ one can use $^4F_{5/2}$ (in Er$^{3+}$:LaF$_3$ $\tau$ = 658 ns [56]).

Let us now consider the use of the Tm$^{3+}$ ion for OQC. This ion provides several choices for the qubit levels and the auxiliary level (see Table 1; the important levels of this ion are shown in Figure 3). For example, similarly to the Pr$^{3+}$ and Er$^{3+}$ ions, the lowest level $|^3H_6\rangle$ of the Tm$^{3+}$ ion has a rather large diagonal element of matrix $U^{(2)}$ ($|U^{(2)}_{H_6H_6}|^2 = 1.25$, see Table 3), and it can be used as an auxiliary level $|1'\rangle$. In this case, levels $|0\rangle = |^3F_4\rangle$ ($E = 5619\,\text{cm}^{-1}$, $|U^{(2)}_{F_4F_4}|^2 = 0.01$, lifetime $\tau$ = 18.05 ms in Tm$^{3+}$:LiYF$_4$ [58]) and $|1\rangle = |^1D_2\rangle$ ($E = 27830\,\text{cm}^{-1}$, $|U^{(2)}_{D_2D_2}|^2 = 0.197$, lifetime $\tau$ = 70 µs in Tm$^{3+}$:LiYF$_4$ [55]) can be used as qubit levels. Corresponding elements of Judd-Ofelt matrices are given in Table 3. The levels of the qubit are separated from other levels of the Tm$^{3+}$ ion below by a noticeable gap exceeding 0.5 eV, which ensures low rates of non-radiative decay of these levels. In this case, the initial state $|0\rangle$ is prepared by applying a pulse with frequency of 12518 cm$^{-1}$ to the allowed ($U^{(2)}_{g0} = 0.237$) transition $|g\rangle \to |0\rangle$ from the ground state $|g\rangle = |^3H_6\rangle$. The large value of $U^{(2)}_{1'1'}$ (4.88) suggests that this scheme may be used for implementation of CNOT gate in case of remarkable mean distance between Tm$^{3+}$ ions.

In the case of the $Tm^{3+}$ ion, high energy level $\left|^{1}I_{6}\right\rangle$ ($E = 34684\,cm^{-1}$, lifetime $\tau = 300\,\mu s$ in $Tm^{3+}$: β-NaYF$_4$ [59] ) has especially large diagonal element of matrix $U^{(2)}$ ($\left|U^{(2)}_{I_6 I_6}\right|^2 = 4.88$, see Table 3), and it can be used also as auxiliary level. Then one can use the levels $|0\rangle = \left|^{3}H_{4}\right\rangle$) and $|1\rangle = \left|^{1}D_{2}\right\rangle$ as qubit levels (see Figure 2). Very large value of $\left|U^{(2)}_{I_6 I_6}\right|^2$ suggests that this scheme may be used for implementation of CNOT gate in case of large mean distance between $Tm^{3+}$ ions.

The values of the matrix elements $U^{(k)}_{01'}$ and $U^{(k)}_{11'}$ given in Table 3 show that the corresponding transitions are sufficiently allowed to use the five light pulses noted in Figure 2. The single-qubit transition $|0\rangle \leftrightarrow |1\rangle$ (not shown in Fig. 1) is also quite well allowed ($U^{(2)}_{01} = 0.127$, $U^{(6)}_{01} = 0.228$), its frequency 15312 cm$^{-1}$ is in convenient spectral range. Note that in case of the auxiliary level $|1'\rangle = \left|^{1}I_{6}\right\rangle$, one can also use the levels $\left|^{3}F_{4}\right\rangle$ (see Table 3) or $\left|^{1}G_{4}\right\rangle$ ($E = 21172\,cm^{-1}$, in $Tm^{3+}$:LiYF$_4$ the lifetime $\tau = 837\,\mu s$ [58]) as one of the qubit levels. In this case, the frequencies of the laser pulses used will be different. Which option should be used depends on the available laser equipment.

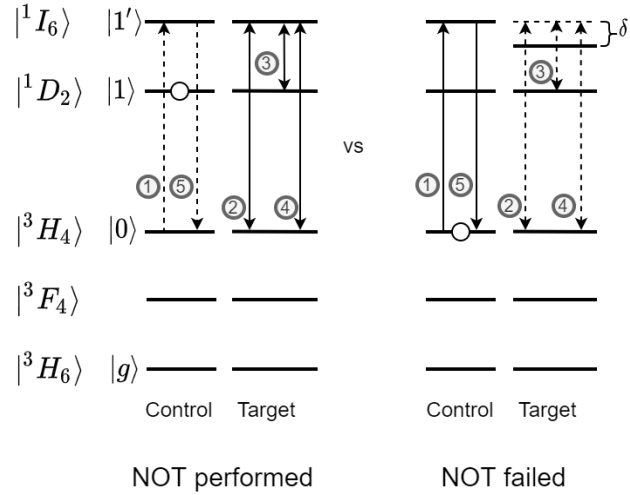

Fig. 2. Main energy levels and CNOT gate operation scheme for $Tm^{3+}$ in the case if the upper state $\left|^{1}I_{6}\right\rangle$ works as auxiliary state $|1'\rangle$. Otherwise, notations are the same as in Figure 1.

Table 3. Frequencies, lifetimes and elements of Judd-Ofelt matrices $U^{(k)}$ of $Tm^{3+}$ levels

| ion | level | $E$, cm$^{-1}$ | $\tau$, $\mu s$ | $U^{(2)}\ U^{(4)}\ U^{(6)}$ | | | | |
|---|---|---|---|---|---|---|---|---|
| | | | | $\left|^{3}H_{6}\right\rangle$ | $\left|^{3}F_{4}\right\rangle$ | $\left|^{3}H_{4}\right\rangle$ | $\left|^{1}D_{2}\right\rangle$ | $\left|^{1}I_{6}\right\rangle$ |
| | $\left|^{3}H_{6}\right\rangle$ | 0 | ∞ | 1.25 .691 .772 | .537 .726 .238 | **.237** .109 .595 | 0 **.316** .093 | .011 .039 .001 |
| | $\left|^{3}F_{4}\right\rangle$ | 5610 | 18050 | **.537** .726 .238 | .001 .409 .269 | .129 .130 .206 | **.575** .096 .023 | 0 .108 0 |
| $Tm^{3+}$ | $\left|^{3}H_{4}\right\rangle$ | 12518 | 2890 | **.237** .109 .595 | .129 .130 .206 | .268 1.62 .583 | **.127** .012 .228 | .066 **.305** .097 |
| | $\left|^{1}D_{2}\right\rangle$ | 27830 | 70 | 0 **.316** .093 | **.575** .096 .023 | **.127** .012 .228 | .197 .004 0 | 0 .051 **.838** |
| | $\left|^{1}I_{6}\right\rangle$ | 34684 | 300 | .011 .039 .001 | 0 **.108** 0 | .066 **.305** .097 | 0 .051 **.838** | 4.88 1.58 .119 |

We emphasize that according to Tables 2 and 3, the lifetime of all levels used in the presented schemes exceeds $10\,\mu s$. In the low temperature limit, this lifetime also determines the decoherence (and phase relaxation) time. Consequently, in the cases considered here, the decoherence time can be more than 10 µs at sufficiently low temperature. Therefore, one can expect that the condition $\Gamma_{inh}/\Gamma_h > 10^5$ required for a fast OQC can actually be met at sufficiently low temperature.

Similarly, CCNOT and other conditional multi-gate operations can be performed by using resonant $\pi$- pulses for $|0\rangle \leftrightarrow |1'\rangle$ transitions in control qubits. Then the NOT operation on the target qubit will be successfully performed if all control qubits are in the state $|1\rangle$.

## 5. Measurement of the final state of qubits

At the end of typical calculations on quantum computers, the main registers are in the half-excited states. To obtain the result of the calculations, the *N*-qubit Hadamard operation is applied to these states [47, 48]. This operation generates the state

$$\left|\Psi^{(f)}\right\rangle = 2^{-N} \prod_m \left(\left(1+e^{i\varphi_m^{(f)}}\right)|0_m\rangle + \left(1-e^{i\varphi_m^{(f)}}\right)|1_m\rangle\right) \qquad (10)$$

with different final phases $\varphi_m^{(f)}$ and population factors $\cos^2\left(\varphi_m^{(f)}/2\right)$ and $\sin^2\left(\varphi_m^{(f)}/2\right)$ of the zeroth and first levels, respectively. One can find these population factors by measuring the optical spectrum of the crystal in the final state: absorption (excitation) spectrum, or/and (stimulated) emission spectrum, or/and Raman spectrum. In the considered case of mixed microcrystals doped with rare-earth ions, the spectrum of these crystals should consist, due to the large inhomogeneous width and the small homogeneous width of ZPLs, of different lines. Intensities and shapes of these lines depend on the population factors. This means that the readout operation can be performed by measuring the entire optical spectrum. This can be done using the method of site-selective spectroscopy of single molecules [60, 61]. To reduce the calculation errors, the calculation and measurement should be repeated several times.

## 6. Conclusion

Here we came to the conclusion that mixed microcrystals like $La_{1-x}Y_xF_3$, $(SrF_2)_x(CaF_2)_{1-x}$, and similar, doped with rare-earth ions ($Tm^{3+}$ and others), can serve as possible physical systems for the fabrication of fast OQC. The qubits in these systems correspond to the quantum levels of the $^4f$ electrons of the rare-earth ions (REI), and they have an optical frequency; an external magnetic field is not required. We found that such a possibility follows from three main properties of REI: weak interaction with the environment, strong inhomogeneous crystal field and the existence of a large number of $^4f$ states with very different values of the reduced elements of the Judd-Ofelt matrices $U^{(2)}$, $U^{(4)}$ and $U^{(6)}$, which determine the properties of these states and the oscillator strengths of $^4f$ - $^4f$ transitions. Most important for the proposed fast OQCs is the ability to simultaneously find both weak (with small diagonal elements of matrix $U^{(2)}$) and strong (with large diagonal elements of this matrix ) interacting two-level systems using these states. For OQC, $^4f$ states with small diagonal elements of the matrix $U_{jj}^{(2)}$ can be used as qubit levels. In contrast, $^4f$ states with large diagonal elements of the matrix $U_{jj}^{(2)}$ can be used as auxiliary levels to implement CNOT and other control gate operations. In this case, the interaction of levels is sufficiently strong to observe the Stark blockade required for conditional gate operations.

We have found that quadrupole-quadrupole interaction is the main cause of the Stark blockade at significant distances between REI. Previously, this was thought to be a dipole-dipole interaction; therefore, the term "dipole blockade" was used in [1-14] to describe the phenomenon.

The important advantage of mixed crystals under consideration is that, despite the ordered crystal lattice, it is possible to have large variations in the local structure. Therefore, in these crystals, inhomogeneous width of ZPL can be very large ($\Gamma_{inh} \sim 1\,\text{THz}$), together with a small homogeneous width $\Gamma_h \ll 0.1\,\text{GHz}$ at low temperatures (which is a consequence of the weak interaction of $^4f$ electrons with phonons and a small number of tunnelling systems and pseudo-local modes of low frequencies). This makes it possible to have $\Gamma_h$ four to seven orders of magnitude smaller than the inhomogeneous width ($\Gamma_h/\Gamma_{inh} \sim 10^{-4}-10^{-7}$) and to meet the conditions of fast OQC $\Gamma_{inh} \gg N\Gamma_L > N\Gamma_h$ for large number $N$ of qubits. In addition, these crystals can have a remarkably high concentration of optical centers, allowing a sufficiently strong interaction of the centers in auxiliary states to be achieved, which is necessary to perform controlled gate operations.

We have also described ensembles of $N$ nearest REIs that can act as an instance of the OQC, and consider using the spectral hole burning method to determine the frequencies of ion qubits in this ensemble.

## Acknowledgments

The work was supported by Estonian Research Council grant PRG347.